\documentclass[%
 aip,
 amsmath,amssymb,
 reprint,%
]{revtex4-1}

\usepackage{graphicx}
\usepackage{dcolumn}
\usepackage{bm}

\usepackage[utf8]{inputenc}
\usepackage[T1]{fontenc}
\usepackage{mathptmx}
\usepackage{etoolbox}
\usepackage{verbatim}
\usepackage{hyperref}
\usepackage{bbold}
\usepackage{xcolor}
\usepackage{mathrsfs}

\usepackage[english]{babel}

\makeatletter
\def\@email#1#2{%
 \endgroup
 \patchcmd{\titleblock@produce}
  {\frontmatter@RRAPformat}
  {\frontmatter@RRAPformat{\produce@RRAP{*#1\href{mailto:#2}{#2}}}\frontmatter@RRAPformat}
  {}{}
}%
\makeatother
\begin{document}

\title{Quantum transport in the presence of a chiral molecular potential}

\author{R. Alhyder}
\email{ragheed.alhyder@ist.ac.at}
\affiliation{Institute of Science and Technology Austria (ISTA), \\
	Am Campus 1, 3400 Klosterneuburg, Austria}


\author{M. Lemeshko}
\affiliation{Institute of Science and Technology Austria (ISTA), \\
	Am Campus 1, 3400 Klosterneuburg, Austria}

\author{A. Cappellaro}
\email{alberto.cappellaro@unipd.it}
\affiliation{Institute of Science and Technology Austria (ISTA), \\
	Am Campus 1, 3400 Klosterneuburg, Austria}
\affiliation{Department of Physics and Astronomy "G. Galilei", University of Padova, \\
    Via Marzolo 8, 35131 Padova, Italy}
\affiliation{National Institute of Nuclear Physics (INFN), Padova Section, \\
    Via Marzolo 8, 35131 Padova, Italy}

\date{\today{}}

\begin{abstract}
We investigate quantum transport in a two-dimensional electron system coupled to a chiral molecular potential, demonstrating how molecular chirality and orientation affect charge and spin transport properties. We propose a minimal model for realizing true chiral symmetry 
breaking on a magnetized surface, with a crucial role played by the tilt angle of
the molecular dipole with respect to the surface. For non-zero tilting, we show that the
Hall response exhibits clear signatures of chirality-induced effects, both in charge and
spin-resolved observables. 
Concerning the former, tilted enantiomers produce asymmetric Hall conductances 
and, even more remarkably, the persistence of this feature 
in the absence of spin-orbit coupling (SOC) signals how 
the enantiospecific charge response results from electron scattering off the 
molecular potential. Concerning spin-resolved
observables where SOC plays a relevant role, 
we reveal that chiral symmetry breaking is crucial in enabling spin-flipping
processes. 
\end{abstract}

\maketitle

\section{Introduction and motivations} 
In recent years, a significant experimental effort has been made to control the static and transport properties
of metallic substrates through the adsorption of chiral molecules
on magnetized or superconducting surfaces
\cite{naaman-2003,yamamoto-2004,ernst-2012,BenDor2013,BenDor2017,Alpern-2019}.
This highly non-trivial interplay between quantum transport, molecular chirality 
and magnetism sits at the core of the so-called Chiral-Induced Spin Selectivity (CISS),
an umbrella term encompassing a wide range of phenomena where spin-dependent observables
and chiral symmetry breaking are intertwined, spanning open questions from
solid-state physics to biochemistry
\cite{xie-2011,gutierrez-2012,guo-2012-1,guo-2012-2,naaman_spintronics_2015,varela-2016,aragones-2017,alam-2017,lu-2019,sessoli-2019,waldeck-2019,geyer-2019,vanwees-2019,volosniev-2020,vanwees-2020,naaman-2020,volosniev-2021,kulkarni-2020,aiello-2022,evers_theory_2022,ozturk-2023,wasielewski-2023,el-naggar-2023}.

Investigating chirality in experimental setups is challenging due to the complex 
interplay between chiral symmetry breaking and intrinsic symmetries of the system's 
observables. As a result, discerning chirality-driven effects is a non-trivial task,
especially on surfaces \cite{ernst-2012,alhyder-2023}. Indeed, chiral symmetry involves 
transforming an object into its mirror image, with such pairs called enantiomers.
For chiral symmetry breaking to manifest in distinct and measurable effects, 
enantiomers must result in non-trivial symmetry breaking mechanisms. This is crucial for understanding how 
chirality affects physical observables, especially for adsorption
experiments. There is indeed an increasing amount of evidence that surfaces
with an out-of-plane magnetization enable an enantiospecific adsorption process
\cite{banerjee-ghosh-2019}, provided the reaction kinetics occurs on a faster timescale
than thermodynamic equilibration.
Within the CISS framework, this observation is understood
in terms of an emergent spin-exchange interaction arising from the interplay between
the magnetized surface and the molecular electric polarizability \cite{fransson-2021}.

Once this enantiospecific process is established, 
the main focus of this paper revolves around the resulting chirality-driven
signature on transport observables, such as, for instance, charge and 
spin-resolved conductances. By leveraging a minimal model presented in 
Sec. II, we are able to show that, in a four-terminal setup, enantiospecific
signals are present both for charge ans spin transport (Sec. III). This occurs in the transverse 
direction with respect to the injected probe current, a phenomenology reminiscent of
the anomalous Hall effect \cite{nagaosa-2010} (AHE). Remarkably, the enantiospecificity
of the AHE charge signal appears to be purely driven by electrons scattering
off the molecular potential; relating to skew-scattering and side-jump mechanisms
\cite{smit1958,berger1970,nagaosa-2010,nagaosa-2018} rather than the intrinsic 
effects \cite{karplus_hall_1954,luttinger1958}, which results from the band structure
geometry. As expected, spin-orbit coupling (SOC) becomes
increasingly relevant for spin observables. 
However, we notice that spin-flipping processes are inhibited whenever
chiral symmetry is not broken and one can single out a mirror transformation 
connecting the two enantiomers on the plane.
We conclude in Sec. IV by commenting how our results can be possibly framed in terms of 
the so-called chirality-induced spin selectivity (CISS) and the impact on the development
a chirality-enhanced spintronic devices \cite{naaman_spintronics_2015,aiello-2022}.


\section{The model.} We examine
 a minimal theoretical framework able to capture 
transport properties across a magnetized two dimensional substrate. 
We consider the following Hamiltonian for electrons
moving in a magnetized substrate,
\begin{equation}
    \hat{H}_{\text{tot}} = \hat{H}_0 + \hat{H}_{\text{SOC}} 
    \label{model hamiltonian}
\end{equation}
where 
\begin{equation}
    \hat{H}_0 = \frac{\hat{\mathbf{p}}^2}{2m^*} + \Delta \sigma_z + V_{\text{dip}}(\mathbf{r}) + V_{\text{ext}}(\mathbf{r})
    \label{eq:H0}
\end{equation}
Here, $\hat{\mathbf{p}}$ is the momentum operator of the electron, $m^*$ is the effective mass of the electron, $\Delta$ is the exchange field resulting from the magnetized substrate. 
$V_{\text{ext}}$ is the confining potential that defines the scattering region.
$V_{\text{dip}}$ is the interaction between electrons and the electric field generated by molecular dipoles.

The molecular potential is modelled as
\begin{equation}
V_{\text{dip}}(\mathbf{r}) = e \big(\bm{\mathcal{E}}_{\mu} \cdot \mathbf{r} \big) 
\exp \bigg\lbrace - \frac{\xi_x^2 x^2 + \xi_y^2 y^2 + \xi_z^2 z^2}{2} \bigg\rbrace,
\label{Vdip potential}
\end{equation}
where $e$ being the electron charge, $\bm{\mathcal{E}}_{\mu}=8\bm{\mu} /\bm{l}^3$ relates to the
electric field proportional to $\bm{\mu} = (\mu_x,\mu_y,\mu_z)^T$ which acts as the molecular 
dipole moment, and $\bm{l}$ is a vector containing the dimensions of the molecule. This form of 
potential has been used for scattering properties before \cite{Ghazaryan2020a}, and is employed
here as a model potential to capture the main aspects of the molecular electric field.

The potential in Eq. \eqref{Vdip potential} breaks chiral symmetry. Indeed, a mirror transformation 
along the $x$ or $y$ axes leads to a rotated potential that is different except 
for $\mu_x = \mu_y$.
Moreover, moving from one enantiomer to the other requires flipping the sign of either $
\mu_x$ or $\mu_y$. For the sake of simplicity, we keep the sign of $\mu_x$ fixed 
and change $\mu_y$, as the parameter allowing us to switch between both chiralities.

In its current form, the effect of the potential on observables is trivial, since changing the sign 
of $\mu_y$ in Eq. \eqref{eq:H0} is equivalent to applying a mirror transformation with respect to 
the $xz$ plane. This leads to any observable having the same value for both enantiomers after 
applying mirror transformation, as we shall see. This is due to the kinetic term 
in $\hat{H}_0$ being invariant under mirror symmetry.

To remedy this problem, and explore how chiral symmetry-breaking affects the whole Hamiltonian, we 
consider the possibility of tilting the molecule with respect to the substrate. 
This is done by introducing an
angle $\theta_w$ togethere with the corresponding rotation matrix 
$\mathcal{R}_{\hat{x}}(\theta_w)$ around the 
$x$-axis. Therefore, positions and electric field vectors in Eq. \eqref{Vdip potential} transform 
into the rotated frame defined as $\mathbf{r}' = \mathcal{R}_{\hat{x}}(\theta_w)\mathbf{r}$, with 
$\mathbf{r}',\mathbf{r}$ 
being the coordinate vectors in the lab, molecular frame respectively. This 
leads to the following expression
\begin{equation}
V_{\text{dip}}^{LR}(\mathbf{r}) =  e \big(\bm{\mathcal{E}'}_{\mu} \cdot \mathbf{r}' \big) 
\exp \bigg\lbrace - \frac{\xi_x^2 x^2 + \xi_y^2 y'^2 + \xi_z^2 z'^2}{2} \bigg\rbrace,
\label{eq:chiral_potential}
\end{equation}
with 
\begin{equation}
\mathbf{\mathcal{E}}'_{\mu} = 8
\begin{pmatrix}
\mu_x/l_x^3 \\
 \\
(\pm \mu_y \cos\theta_w - \mu_z \sin\theta_w) / l_y^3 \\
 \\
(\pm \mu_y \sin\theta_w + \mu_z \cos\theta_w)/l_z^3
\end{pmatrix}
\end{equation}
and
$\mathbf{r}' = (x, y \cos\theta_w - z\sin\theta_w, y \sin\theta_w + z \cos\theta_w )$. Different
signs of $\mu_y$ corresponds to the enantiomers. Note that flipping the sign of $\mu_y$ here is 
not equivalent to applying a mirror transformation, providing true chiral symmetry breaking in $
\hat{H}_0$, and this point is crucial when considering chirality-dependent effects.
%
%
Consequently, in the rotated frame, changing the sign of $\mu_y$ is sufficient to switching between 
opposite enantiomers for $\theta_w \neq 0$, without it being equivalent to mirror transformation.
Within our model, the potential experienced by electrons on the surface effectively 
depends on the handedness of the adsorbed molecule only if $\theta_w \neq 0$. If that is
not the case, the L- and R-potentials in Eq. \eqref{Vdip potential} are mirror-symmetric.
As we show in the following, this implies that, when considering different enantiomers, 
the response is trivial for $\theta_w = 0$. When the molecule is tilted
(i.e. $\theta_w \neq 0$ in Eq. \eqref{Vdip potential}), the charge and spin response strongly
depend on the potential handedness. An analogous scenario is actually observed in the 
enantiospecific adsorption of chiral molecules upon out-of-plane magnetized surfaces
\cite{banerjee-ghosh-2019}. The enantiospecificity of the molecule-surface 
interaction is rationalized in terms of an emergent selective spin-exchange interaction and 
its energetics. More precisely, the substrate is inducing an electric dipole polarization, which is
in turn accompanied by a spin polarization whose orientation depends on the molecular chirality 
\cite{fransson-2021}. Thus, assuming the presence of this exchange interaction with the
magnetized surface, it is possible to extract an effective potential for both enantiomers;
according to density-functional calculations \cite{banerjee-ghosh-2019}, 
the energy separation between their corresponding
minima lies well above the scale of room-temperature fluctuations. 
It also important
to remark that, as mentioned in the introduction, in order to observe this 
enantiospecific process, the reaction timescale has to be shorter than the 
one related to thermal equilibrium. This feature is strongly dependent from
the particular molecular species and experimental conditions, appearing (for now)
beyond the grasp of a reliable theoretical modelling \cite{banerjee-ghosh-2019}.
Nevertheless, while we do not claim to reproduce 
this dynamical and molecule-dependent process, 
our minimal model actually includes two of the crucial ingredients leading
to the above described enantioselective process, i.e. the molecular electric polarizability and the presence of a magnetic substrate.

As an additional remark, while
it is common to distinguish different enantiomers by referring to their handedness 
(i.e. left vs. right), within our framework this denomination
is purely conventional, since these labels do not reflect the geometric 
structure of the molecule, but rather the different nature of the two potentials considered 
\cite{doyle-2013}.
Finally, the field associated with $V_{\textrm{dip}}(\mathbf{r})$ as in 
Eq. \eqref{Vdip potential} is responsible for the spin-orbit interaction modelled on 
the Rashba model \cite{rashba-2003,rashba-2004}, leading to an additional term defined as 
\begin{equation}
\hat{H}_{\text{SOC}} = - (\alpha_R /\hbar)\; \bm{\sigma} \cdot 
\big[\mathbf{E}_{\text{dip}}(\mathbf{r}) \times \hat{\mathbf{p}} \big]
\label{rashba soc}
\end{equation}
where $\bm{\sigma}$ is a vector made of the usual Pauli matrices, 
$\mathbf{E} = - \bm{\nabla} V_{\text{dip}}(\mathbf{r})$, with $V_{\text{dip}}$ being the chiral 
potential and $\alpha_R$ is the Rashba spin-orbit coupling parameter. Note that all spin-
independent observables trace out the SOC contribution, and the discussion above still holds. 
However, spin-dependent observables will have contributions coming from the SOC, which turn out to 
be chirality-dependent, 
as shall be explored in the following sections.

\begin{figure}[ht!]
\centering
\includegraphics[width=0.85\linewidth]{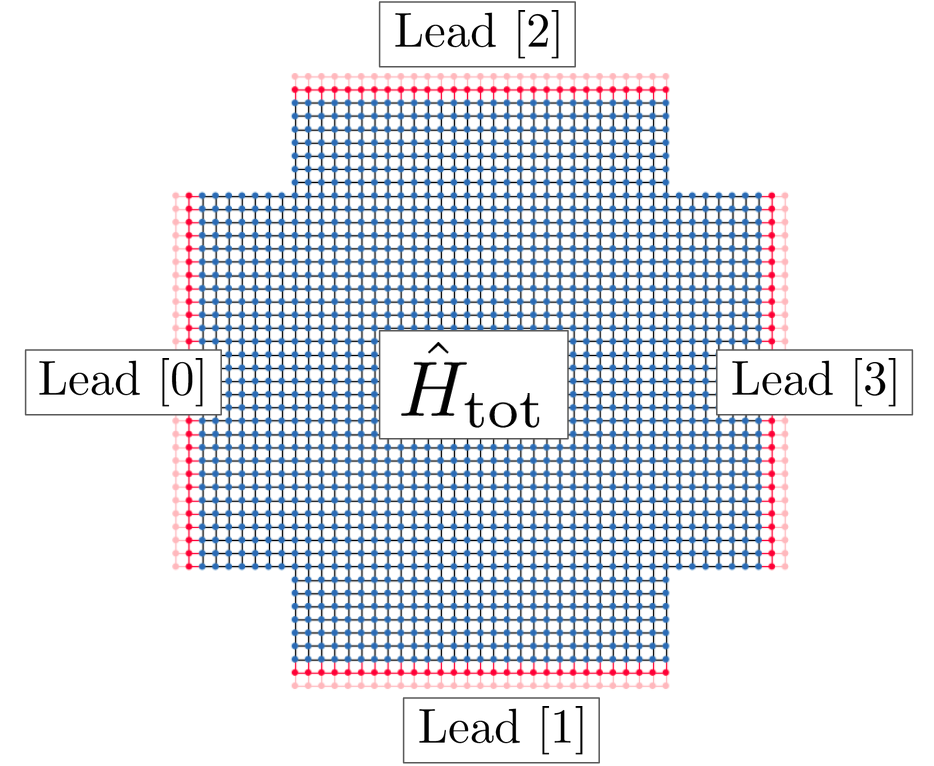}
\caption{Instance of a four-terminal Hall setup. 
The Hamiltonian $\hat{H}_{\textrm{tot}}$, as given by Eq. \eqref{model hamiltonian}, 
is implemented in the central (blue) region. 
Four semi-infinite leads (red) are attached to it. In order to lessen the onset of
shape-induced resonances, in the numerical simulations we actually consider a square-shaped
scattering region, whose dimension is $56 \times 56$ sites.
}
\label{fig:1}
\end{figure}

Moving to the central issue of chiral symmetry breaking and its signature on
transport observables, here
we consider a four-terminal device as shown in Fig. \ref{fig:1}. This implements a so-called
Hall bridge, where $\hat{H}_{\textrm{tot}}$ as in Eq. \eqref{model hamiltonian} acts on
the central region, specified by blue points in the above mentioned figure. Four semi-infinite
leads are attached to it, acting as electrodes in standard transport experiments. These leads can be thought of as waveguides, driving plane waves
in and out of the scattering region. For this reason, the leads Hamiltonian is written as 
$\hat{H}_L = \mathbf{k}^2/2m^* + \Delta \sigma_z$ in momentum representation, 
such that spin-up/spin-down degeneracy 
is already lifted.

In order to compute transport observables, we rely upon the scattering 
wavefunctions formalism \cite{datta-book-1995}. While it is completely 
equivalent to the Keldysh method based on non-equilibrium Green's functions
\cite{gaury-2014}, it is more convenient from a computational perspective and has
been implemented in our simulation through the Kwant package \cite{Groth-2014,Kloss-2021}.
This method relies on casting the Hamiltonian $\hat{H}_{\text{tot}}$ on a tight-binding discretized model with hopping parameters $t = \hbar^2/(2m^* a^2)$, where $a$ is the lattice constant,
Now, considering the four-terminal device displayed in Fig. \ref{fig:1},
when a current $I_0$ is inserted in the left lead, we can measure the (anomalous)
Hall response by looking at the voltage drop in the transverse direction, i.e. between the 
top and bottom leads, as $V_H = V_2 - V_1$. Lead voltages are simply extracted by 
inverting the Ohm's law $\hat{\mathbf{G}} \cdot \mathbf{V} = \mathbf{I}$, where $\hat{\mathbf{G}}$
is the conductance matrix. Its elements $G_{ij}$ represent the conductance between lead $i$ and $j$ ($ i\neq j$) as computed by solving numerically the scattering problem. 
More precisely, for a fixed energy $E$, we have \cite{datta-book-1995,schliemann-2009}
\begin{equation}
G_{ij}(E) = \bigg( \frac{e^2}{h} \bigg)\sum_{n\in i} \sum_{n' \in j} |t^{ij}_{nn'}(E)|^2
\label{conductance landauer formalism transmission}
\end{equation}
where $n$ is the index labelling the modes open for conduction at energy $E$ in lead $i$, and 
similarly for $n'$, with $t^{ij}_{nn'}(E)$ the corresponding transmission scattering amplitude. 
More specifically, these coefficients are the elements of $\hat{\mathbf{t}}^{ij}(E)$,
a matrix whose dimension is $\mathcal{N}^i(E) \times \mathcal{N}^j(E)$, with $\mathcal{N}^{i(j)}(E)$
is again the number of modes open conducting modes in lead $i$ ($j$). In turn, 
$\hat{\mathbf{t}}^{ij}(E)$ is an off-diagonal block of 
the whole scattering matrix $\hat{\mathbf{S}}(E)$ 
for the $N$-terminal device. 

Except when otherwise specified, 
the system is solved using experimentally relevant parameters, where
we set the effective mass and the lattice 
constant, $m^* \simeq 0.067 \; m_e$ and $a\simeq 7.1 \; \text{\AA}$; energies
are reported in units of the hopping parameter $t = \hbar^2 /(2m^* a^2)$,
which is readily computed to be around $1.14$ eV. 
As for the Rashba SOC, we set $\alpha_R = 0.1\; \text{eV} \cdot \text{\AA}$; corresponding, 
in our simulations, to $\alpha_R / (2 a) \simeq 0.006 \; t$. 
In order to extract $V_H$, we apply a current $I_0 = 3\; \text{e}\cdot t/h \simeq 1.0\; \text{mA}$
onto the left lead, with respect to Fig. \ref{fig:1}.
As mentioned in the previous section, spin degeneracy is removed by considering the 
exchange field coming from the magnetized surface, here $\Delta \simeq 0.001\; t$.
As for the molecular potential $V_{\text{dip}}(\mathbf{r})$
in Eq. \eqref{Vdip potential}, our values are comparable
to the 1,2-propanediol molecule, such that $l_x = 2.0$ nm, $l_y = 4.0$ nm and $l_z = 10$ nm, while
the electric dipole components are set to $\mu_x = 2.4$ D, $\mu_y = 5.0$ D and $\mu_z = 1.8$ D
\cite{lovas-2009}.

\section{results}

\subsection{Conductance and Hall voltage}

\begin{figure}
    \centering
    \includegraphics[width=0.99\columnwidth]{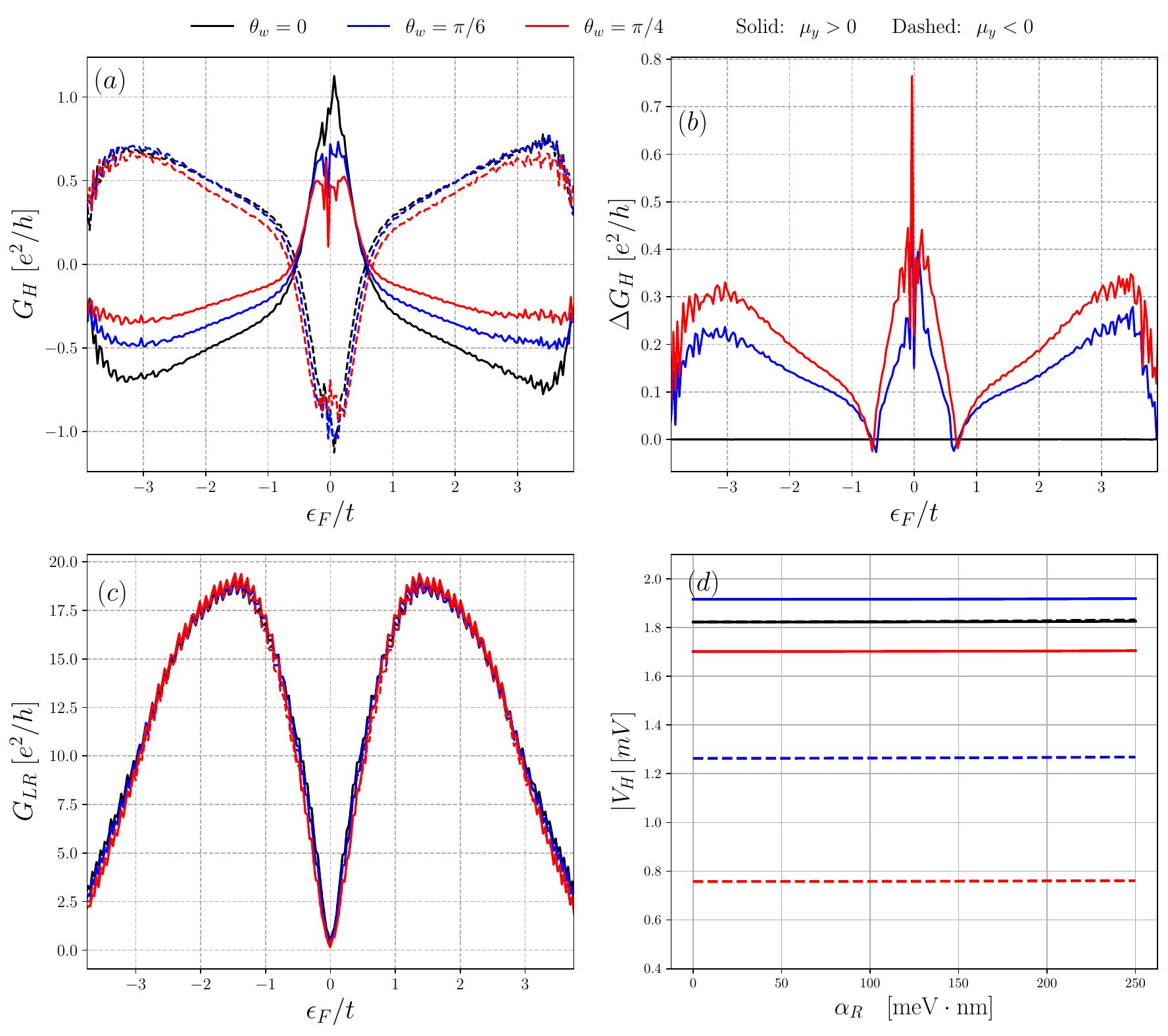}
    \vspace{-10pt}
    \caption{
    (a). Hall conductance $G_H$ (defined in Eq. \eqref{hall-conductance definition}) 
    as a function of the Fermi energy 
    $\epsilon_F$, here in units of the hopping parameter for the tight-binding model 
    $t = \hbar^2 / (2m^* a^2)$. As pointed out in the main text, 
    $\alpha_R / (2a) \simeq 0.006 \, t$ and $\Delta \simeq 0.001\, t$. We report three different values 
    of the tilting angle $\theta_w$, from $\pi/4$ (red lines) down to $\pi/6$ (blue lines) and $0$ (black 
    lines), for both of potential's handedness, with solid lines corresponding to $\mu_y > 0$ and
    dashed ones to $\mu_y < 0$. 
    (b). Plot of $\Delta G_H = |G_H^{(L)}| - |G_H^{(R)}|$ as a function of the Fermi energy
    for the same three values of $\theta_w$ listed above.
   (c). Longitudinal conductance, defined as $G_{LR}(E) = G_{30}(E)$, where the 
    subscripts refer to Fig. \ref{fig:1} lead numbering. As for the above panels, $\theta_w = 0,\, \pi/6$
    and $\pi/4$, but here no difference is observed due to the potential chirality.
    (d). Absolute value of the Hall voltage $V_H$ at $\epsilon_f /t \simeq 0.1$
for different values of the spin-orbit coupling constant $\alpha_R$,
while the rest of simulation parameters are the same as in
the previous figures. For the sake of clarity we just consider
the case of $\theta_w = 0$ (no  difference between the two
enantiomers) and  $\theta_w = \pi/4$ , where we see a spread of the
order $\sim\; 1\,\mathrm{mV}$.}
    \label{fig:4panels}
    \end{figure}

First, it is important to explore the dependence of the overall Hall response at different values in such  of the energy and its dependence on SOC strength, as shown in Fig. \ref{fig:4panels}.

We focus on the Hall conductance $G_H$ defined as \cite{schliemann-2009} 
\begin{equation}
G_H(E) = \frac{1}{2} \bigg[G_{20}(E) - G_{10}(E)\bigg]    \;,
\label{hall-conductance definition}
\end{equation}
with the numbering referring to the setup displayed in Fig. \ref{fig:1}. In  Fig. \ref{fig:4panels}.
(a), we report the Hall conductance $G_H$ as a function of the Fermi
energy $\epsilon_F$ for the whole bandwidth $[-4t -\Delta , 4t + \Delta]$ for different values of 
the tilting angle  $\theta_w$. For every value of $\theta_w$, we plot the results for both 
handedness values, with solid lines  corresponding to $\mu_y > 0$ and dashed one to $\mu_y < 0$. It 
is worth noting that, in practice,
we work in the low-temperature limit $ T \ll T_F$, with the Fermi energy determining
how many lead modes are open for conduction. 

By increasing the electron energy from the left band edge (approximately $\sim -4t$), 
and except for 
oscillations resulting from the discretizing procedure and the finite system size, the 
Hall conductance displays a non-monotonic behaviour,
reaching its maximum as we approach to the band center ($\epsilon_F = 0$), and dropping to zero at 
the band edges. The behavior of the conductance is not considerably affected by the molecular 
configuration. However, it is immediate to realize that different enantiomers induce
responses with opposite sign, a feature observed in a series of seminal experiments with absorbed 
chiral molecules upon magnetized surfaces \cite{BenDor2017}. 

More remarkably, our framework also adds another crucial 
detail: depending on the molecular tilting angle $\theta_w$, different enantiomers lead to 
asymmetric responses (i.e. different absolute value for $G_H$). For $\theta=0$, we show that 
changing handedness is equivalent to taking the mirror transformation counterpart of the 
conductance, with no change in the amplitude, as discussed before. 
On the other hand, for $\theta\neq 0$, 
the conductance is non-trivially affected by the symmetry breaking introduced 
in the scattering region by the chiral transformation on the potential. This leads to 
a relevant difference in the amplitudes of $G_H$ for different enantiomers, as evident by 
looking at 
Fig. \ref{fig:4panels}.(b), where we plot 
$\Delta G_H(E) = |G^{(L)}_H| - |G^{(R)}_H|$ for each value of
$\theta_w$ mentioned above. 
There, we see that the response is perfectly symmetric if the potential
is orthogonal to the magnetized substrate ($\theta_w = 0$), but as soon as the molecule is tilted
a difference between left and right enantiomers emerges.
This goes back to the fact that the sign flip in the case of $\theta_w = 0$ is expected, since changing the sign of $\mu_y$ is equivalent to applying a mirror transformation along the y-axis in the scattering region of the molecule.

In order to highlight the peculiar behaviour of $G_H$, we also show that the longitudinal 
conductance $G_{LR}$ between leads $0$ and $3$ in  Fig. \ref{fig:4panels}.(c).  This conductance is 
equally symmetric with respect to $\epsilon_F$ and reaches a maximum around $\epsilon_F/t \approx 
1$ and vanishes at the edges of the bands as expected \cite{Jian2005}. Importantly, we notice how 
the handedness of the superimposed potential does not influence transport in the longitudinal 
direction.

Finally, an experimentally relevant observable is the transverse Hall voltage $V_H$ with respect to 
the spin-orbit coupling, reported in Fig. \ref{fig:4panels}.(d). The results of the numerical 
simulation show a constant value of the Hall voltage $V_H$ 
(at the fixed value of $\epsilon_F / t \simeq 0.1$) 
with respect to the SOC coupling parameter $\alpha_R$. 
The Hall voltage \textit{only} only changes signs by moving 
from one enantiomer to the other one (by flipping the sign of $\mu_y$) 
when the molecule is not tilted ($\theta_w = 0$). 
The reason is once again to be found in the transformation leading to the other enantiomer being 
 equivalent to a mirror transformation 
with repect to the $xz$ plane, flipping in turn the sign of $V_H$. 
Therefore, when plotting the absolute value, 
both enantiomers produce the same values of the Hall voltage. 

On the other hand, for ($\theta \neq 0$), we find that the value of the Hall voltage changes as well as its sign. This behavior mimics the one observed experimentally for the Anomalous Hall effect when chiral molecules are adsorbed on a two-dimensional metal in \cite{BenDor2017}, and while the difference in the absolute value of the resistance was attributed to experimental imperfections, we show that breaking chiral symmetry in the way we outlined 
can lead to such an effect.

It is also important to notice that 
we report the same values of the Hall voltage for $\alpha_R = 0$. This leads to 
the conclusion that the finite value of the voltage is related to electrons scattering off the 
molecular potential, with no need to include additional forces. 
This effect can be related to 
skew-scattering \cite{smit1958} and side-jump \cite{berger1970} contributions to the anomalous Hall 
conductivity.

%
%
%

\subsection{Density of states}
\begin{figure}
    \centering
    \includegraphics[width=0.99\columnwidth]{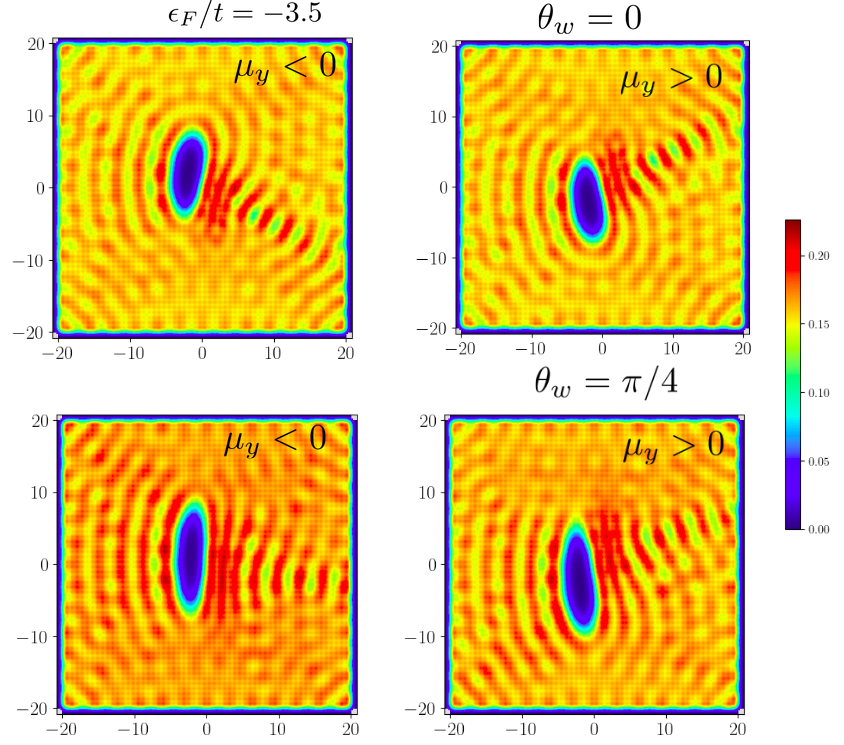}
    \caption{
    Local density of states at $E = -3.5\ t$, as defined in Eq. \eqref{ldos definition}, 
    for both potential handedness ($\mu_y \lessgtr 0$). 
    We consider two tilting angles, $\theta_w = 0$ (top panels) and $\theta_w = \pi/4$
    (bottom panels). Similarly to the choice made in Fig. \ref{fig:4panels},  
    we set $\alpha_R = 0.1\; \mathrm{eV\cdot nm}$, resulting in 
    $\alpha_R / (2a) \simeq 0.006\; \mathrm{t}$, and $\Delta = 0.001\;\mathrm{t}$.
    As discussed in the main text, for $\theta_w = 0$ we can easily identify a mirror 
    plane at $x = 0$, while this is not at all possible when the molecular potential is tilted
    as in the bottom panels.
    }
    \label{fig:ldos-plots}
\end{figure}
In Fig. \ref{fig:ldos-plots} presents the local density of scattering states (LDOS) within 
our four-terminal device. 
It is important to distinguish the LDOS from the conventional density of states (DOS), as the LDOS 
provides spatially resolved information about the distribution of electronic states within the 
scattering region. Formally, the LDOS is simply defined as
\begin{equation}
\rho(\mathbf{r}, E) = \sum_j |\psi^{(S)}_{j}(\mathbf{r})|^2 \; \delta(E - \epsilon_j),
\label{ldos definition}
\end{equation}
where $\psi^{(S)}{j}(\mathbf{r})$ represents the scattering wavefunction at position $\mathbf{r}$, 
and $\epsilon_j$ are the corresponding eigenenergies.

Examining the top panels of Fig. \ref{fig:ldos-plots} for $\theta_w = 0$, we observe that the LDOS 
remains symmetric with respect to the two potential handedness configurations. This symmetry 
suggests that, in this particular setup, changing the chirality of the molecule does not induce a 
significant spatial redistribution of scattering states. The scattering processes in these panels 
are related by mirror symmetry, consistent with the symmetry of the potential for $\theta_w = 0$.

However, when the molecule is tilted ($\theta_w \neq 0$), as shown in lower left panels, this mirror symmetry is broken. The scattering processes in these panels are no longer related by a simple reflection, indicating that the chiral potential asymmetrically influences the spatial distribution of the scattering states. This asymmetry highlights the role of $\theta_w$ in determining the symmetry properties of the scattering processes, with the tilted configuration introducing a chiral-dependent modification to the system’s electronic behavior.

While the LDOS provides valuable insights into the spatial redistribution of states, it remains spin-independent. To gain a deeper understanding of the role of chirality in spin transport, we now turn to spin-resolved quantities, which reveal critical information about spin-selective effects that cannot be captured by the LDOS alone. These spin-resolved measurements are essential for accurately characterizing chirality-induced transport phenomena, as spin-independent observations may overlook important asymmetries arising from spin-dependent interactions.

\subsection{Spin-dependent transport}
To elucidate the importance of chiral symmetry breaking in the problem, we move to spin-dependent observables.
\begin{figure}
\centering
\includegraphics[width=\columnwidth]{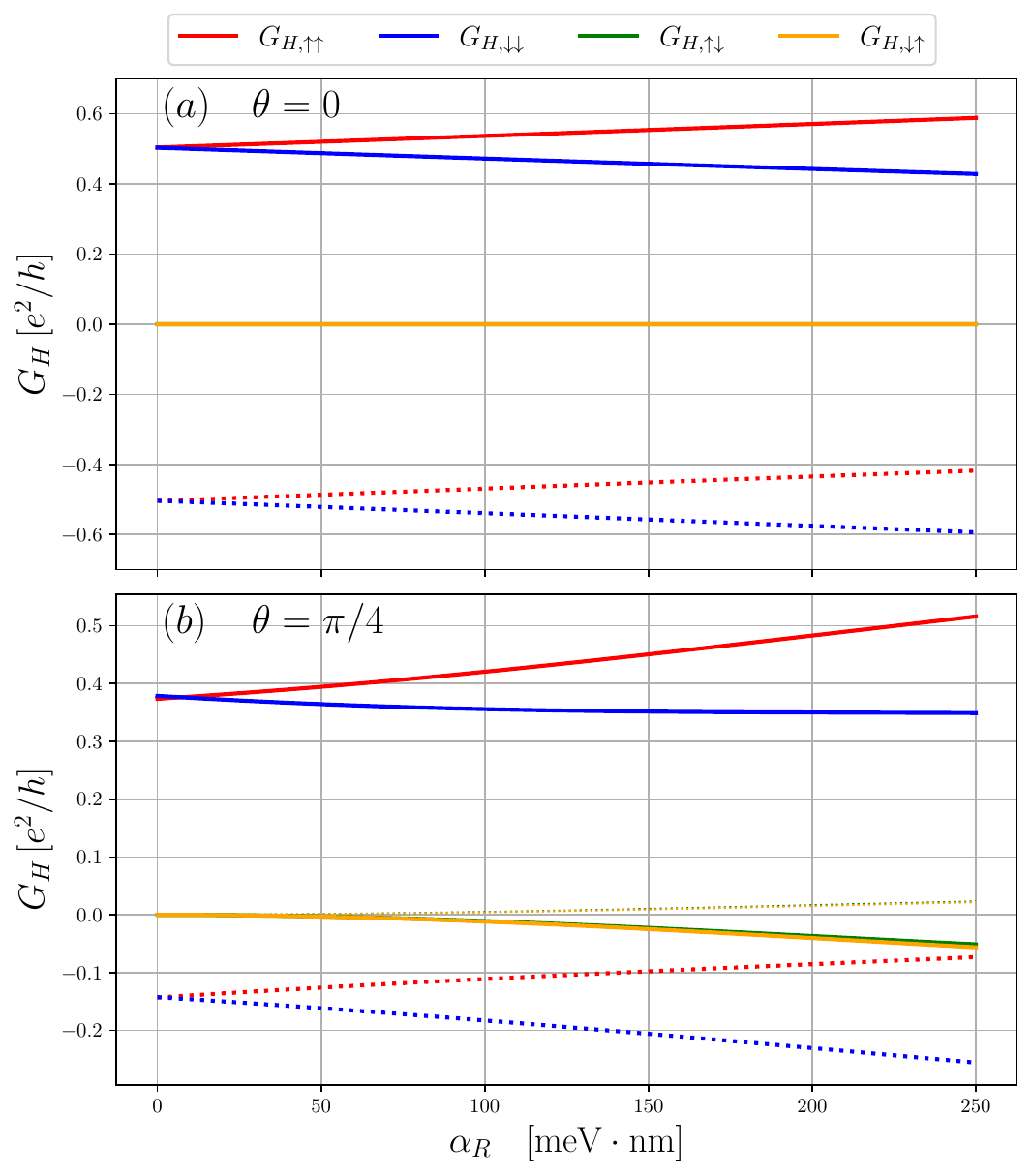}
\caption{Spin dependent Hall conductance for different spin components with $\theta=0$ in panel (a) and $\theta= \pi/4$ in panel (b) as a function of the SOC parameter $\alpha_R$ close to the
band center.
Solid lines are used for $\mu_y >0$ and dashed lines for $\mu_y <0$. 
While $\alpha_R$ is increased up to 
$250\; \mathrm{meV\cdot nm}$, all the other parameters are kept as in Fig. \ref{fig:4panels}
and Fig. \ref{fig:ldos-plots}. Once again, we remark the symmetry observed for $\theta_w = 0$
together with the spin-flipping components of the conductance being strictly zero, 
signalling that these processes are inhibited for this potential mirror-symmetric configuration. 
This significantly changes when the potential is tilted, as discussed in the main text.
}
\label{fig:spin-dependent-cond}
\end{figure}
The spin-resolved Hall conductance curves shown in Fig. \ref{fig:spin-dependent-cond} reveal 
distinct behaviors depending on the orientation angle $\theta$ of the chiral molecule relative to the electronic system. 

For $\theta = 0$ in Fig. (\ref{fig:spin-dependent-cond}.a), the Hall conductance components, $G_{\uparrow\uparrow}$ and $G_{\downarrow\downarrow}$, increase (decrease) linearly, keeping the total conductance constant. In addition, these two components exhibit mirror symmetry. In a consistent manner with the discussion in the spin-independent observables, this symmetry arises from the fact that changing the handedness of the chiral molecule in this configuration is equivalent to applying a mirror transformation, effectively interchanging the spin-up and spin-down components. As a direct consequence, the spin-mixed conductances, $G_{\uparrow\downarrow}$ and $G_{\downarrow\uparrow}$, remain strictly zero, which is consistent with expectations for a system governed by a mirror-symmetric operator. This confirms that, in this regime, spin-flip processes are inhibited, preserving spin coherence in the transport channel.

In contrast, for $\theta = \pi/4$ in Fig. (\ref{fig:spin-dependent-cond}.b), the system no longer exhibits mirror symmetry. While the total conductance remains independent of the SOC parameter 
$\alpha_R$ (cfr. Fig. \ref{fig:4panels}.d), a notable change occurs in the spin-mixed conductance components $G_{\uparrow\downarrow}$ and $G_{\downarrow\uparrow}$. These values deviate from zero and do not mirror each other, indicating that spin-flip transitions are now allowed. The breaking of mirror symmetry by the chiral potential in this tilted configuration thus modifies the spin-dependent scattering processes, enabling coupling between spin-up and spin-down states.

This is an indication that the symmetry-breaking induced by the chiral molecule’s tilted orientation can significantly modify spin-dependent scattering processes, allowing spin-flipping transitions. The observed behavior underscores the critical role of the molecular orientation and chirality in controlling spin transport, highlighting a mechanism through which spin-selective conductance can be engineered.
In the context of the ongoing debate within the CISS community, our results
in Fig. \ref{fig:spin-dependent-cond} suggest 
that the adsorbed molecular structure is actually acting as a spin-polarizer. Indeed, the
RSOI enabling spin-flipping scattering processes, whose enantiospecificity is evident only
when $\theta_w \neq 0$, consistently with our results on charge transport. As previously mentioned,
our minimal model cannot reproduce the full picture of the ongoing spin-polarization (and
in particular its magnitude \cite{gutierrez-2012}). Nevertheless, within a range of reasonable 
experimental values for $\alpha_R$, enantiospecific spin-flipping processes are actually contributing significantly to the whole transport picture.

\section{Conclusions and future perspectives.}
We have explored numerically the effects of breaking chiral symmetry on transport properties of a conventional two dimensional electronic gas. We showed that a significant asymmetry in the hall response arises due to this symmetry breaking. This paves the way for more careful studies of the effects of chiral symmetry breaking on the topological properties of the bands in such system and how this ties in the picture of CISS \cite{naaman_spintronics_2015,evers_theory_2022}. 

We also showed the importance of chiral symmetry breaking on spin-dependent observables, where chirality, if introduced properly into the system, can introduce spin-flip processes. This behavior highlights the profound impact of the molecular orientation and chirality on spin transport. The ability to selectively break spin symmetry through a controllable parameter introduces a powerful mechanism for engineering spintronic devices. In particular, the observed asymmetry in spin-mixed conductances suggests potential applications in spin-selective filtering and spin-based logic operations, where controlling spin-flip processes is crucial. 

Moreover, the persistence of the total conductance’s invariance with respect to spin-orbit coupling underscores that the chiral potential primarily affects spin mixing rather than overall charge transport, emphasizing the role of chirality in tailoring spin responses in low-dimensional systems. 
More importantly, this study highlights the importance of spin-resolved measurements when exploring chiral-induced spin properties.

\section*{Acknowledgements.} 
We thank Artem Volosniev, Narcis Avarvari, Georgios Koutentakis, Sandro Wimberger and Binghai Yan for useful discussions.
R. A. received funding from the Austrian Academy of Science ÖWA grant No. PR1029OEAW03.
 M.L. acknowledges support by the European Research
Council (ERC) Starting Grant No.801770 (ANGULON).
A.~C. received funding from the European Union’s
Horizon Europe research and innovation program under the
Marie Skłodowska-Curie grant agreement No. 101062862 - NeqMolRot.
\section*{Data Availability Statement}
The code to reproduce our results within the scattering wavefunction approach is available upon 
request, and it is based upon the open source Python package Kwant presented in Ref.
\onlinecite{Groth-2014}.
\bibliography{main.bib}


\end{document}